\documentclass[12pt]{article}
\usepackage{epsfig}
\usepackage{amsfonts}

\newcommand{\eq}{\begin{equation}}
\newcommand{\eqx}{\end{equation}}
\newcommand{\eqn}{\begin{eqnarray}}
\newcommand{\eqnx}{\end{eqnarray}}
\newcommand{\f}[2]{\frac{#1}{#2}}

\newcommand{\lm}{\lambda}
\renewcommand{\th}{\theta}
\newcommand{\al}{\alpha}
\newcommand{\bt}{\beta}
\newcommand{\om}{\omega}
\newcommand{\dl}{\delta}
\newcommand{\gm}{\gamma}

\newcommand{\PP}{{\cal P}}
\newcommand{\QQ}{{\mathfrak Q}}
\renewcommand{\SS}{{\mathfrak S}}
\newcommand{\RR}{{\mathfrak R}}
\newcommand{\LL}{{\mathfrak L}}
\newcommand{\HH}{{\mathfrak H}}
\newcommand{\flip}{P}
\newcommand{\cas}{{\cal C}_{12}}
\newcommand{\proj}{\mbox{\rm proj}}
\newcommand{\nn}{{\cal N}}

\newcommand{\qqqq}{\quad\quad\quad\quad}
\newcommand{\qqq}{\quad\quad}

\newcommand{\ket}[1]{\left|{#1}\right\rangle}
\newcommand{\bra}[1]{\left\langle{#1}\right|}

\newcommand{\ant}{{\cal S}}
\newcommand{\chrg}{{\cal C}}

\title{The $AdS_5\times S^5$ superstring worldsheet S-matrix and
  crossing symmetry} 

\author{Romuald A. Janik\thanks{e-mail: {\tt ufrjanik@if.uj.edu.pl}} \\ \\
Institute of Physics\\
Jagellonian University,\\
ul. Reymonta 4, \\
30-059 Krak{\'o}w\\
Poland
}

\begin{document}

\maketitle

\begin{abstract}
An S-matrix satisying the Yang-Baxter equation with symmetries
relevant to the $AdS_5 \times S^5$ superstring has recently been
determined up to an unknown scalar factor. Such scalar factors are
typically fixed using crossing relations, however due to the lack of
conventional relativistic invariance, in this case its determination
remained an open problem. 

In this paper we propose an algebraic way to implement crossing
relations for the $AdS_5 \times S^5$ superstring worldsheet
S-matrix. We base our construction on a Hopf-algebraic formulation of
crossing in terms of the antipode and introduce generalized rapidities
living on the universal cover of the parameter space which is
constructed through an auxillary, coupling constant dependent,
elliptic curve. We determine the crossing transformation and write
functional equations for the scalar factor of the S-matrix in the
generalized rapidity plane.
\end{abstract}

\vfill

\pagebreak

\section{Introduction}

One of the most fascinating discoveries in recent years was the
unravelling of integrable structures in planar $\nn=4$
SYM~\cite{Minahan:2002ve,Beisert:2003tq,Beisert:2003yb,kor1} and in
superstring theory on $AdS_5\times
S^5$~\cite{Bena:2003wd,Arutyunov:2003uj,Dolan:2003uh,Kazakov:2004qf,
Arutyunov:2004yx,Beisert:2005bm,Alday:2005gi}. In view of the AdS/CFT
correspondecne \cite{Maldacena:1997re} which links the two theories
and identifies anomalous dimensions of the gauge theory with energies
of string excitations in $AdS_5 \times S^5$, these discoveries have  
thus opened up the possibility of detailed testing of
the proposed correspondence. Even more interestingly, the techniques of
integrability allow in principle for an exact quantization of the
superstring in $AdS_5 \times S^5$ and/or the determination of anomalous
dimensions in the $\nn=4$ SYM. 

However it turned out that at three loop order there is a set of
discrepancies between gauge theory calculations and string theory
results~\cite{Callan:2003xr,Frolov:2002av,Callan:2004uv,Serban:2004jf}.
This does not signify a contradiction since the domains of
applicability are nonoverlapping and in order to perform a comparision
an extrapolation is needed -- in fact an order of limits problem was
suggested to be a possible explanation. One source of the discrepancy,
the so called `wrapping interactions' have apparently been ruled out
\cite{us,hubbard} so the problem remains open. A key ingredient which
enters the calculation of the anomalous dimensions or string energies
is the S-matrix. `Phenomenologically'
one can quantify the disagreement by a {\em scalar} dressing factor
between the `string' S-matrix and the asymptotic `gauge-theory'
S-matrix \cite{BDS}. Initially this has been proposed to hold in subsectors of
the full theory \cite{dressing} but subsequently it has been
extended to the whole S-matrix. The structure of the dressing factor
on the string theory side appears to be quite complicated with
$1/\sqrt{\lm}$ deviations \cite{28,29,30} from the strong coupling
expression \cite{dressing}.

In fact, in a remarkable series of papers \cite{stS,stBS} the S-matrix
in various sectors was uncovered. This culminated with
\cite{beisert22} where the S-matrix with
$su(2|2)\times su(2|2) \subset psu(2,2|4)$ symmetry was determined up
to an unknown scalar function $S_0(p_1,p_2)$:
\eq
\label{e.sfull}
S(p_1,p_2)=S_0(p_1,p_2) \cdot \left[ \hat{S}_{su(2|2)}(p_1,p_2) \otimes
\hat{S}_{su(2|2)}(p_1,p_2) \right]
\eqx
with $\hat{S}_{su(2|2)}(p_1,p_2)$ being the S-matrix determined
uniquely by $su(2|2)$ symmetry in \cite{beisert22}. A key remaining
problem is the determination of the scalar `dressing factor'
$S_0(p_1,p_2)$. The aim of this paper is to propose functional equations
which $S_0(p_1,p_2)$ has to satisfy.

A similar situation exists in {\em relativistic} integrable quantum
field theories, where symmetries (including the nonlocal Yangian or
affine quantum algebra) determine completely the matrix form of the
S-matrix up to an overall scalar factor. This scalar factor is in turn
determined by requiring unitarity and crossing symmetry of the
S-matrix. Only these two conditions together determine uniquely a
scalar factor with the minimal number of poles/zeroes in the physical
region. The only remaining ambiguities are the CDD factors which serve
to introduce further poles, if needed on physical grounds, but generically
the minimal solution suffices (see \cite{zamolodchikov}).

It is important to emphasize that the two conditions -- unitarity and
crossing have quite a different status. Unitarity is, as is the
Yang-Baxter equation, a consistency condition for the
Faddeev-Zamolodchikov algebra, and does not involve any dynamical
assumptions about the theory in question. The matrix form of the
S-matrix from \cite{beisert22} indeed directly satisfies the
unitarity condition. On the other hand the crossing condition involves
a link between the scattering of particles and the same process with a
particle changed into an anti-particle. The definite form of the
resulting equation involves in an essential way explicit relativistic
kinematics as formulated in the rapidity plane.
In this form it depends crucially on the relativistic invariance of the theory.

It is this last property which makes it very difficult to adopt a
similar strategy in the case of the worldsheet theory of the $AdS_5
\times S^5$ superstring. There the light-cone quantization of the
worldsheet theory, which seems necessary in order to deal
only with physical excitations, is {\em inconsistent} with Minkowski
metric on the
worldsheet~\cite{LCSFT,Callan:2003xr,Callan:2004uv,Arutyunov:2004yx}.
Therefore the guiding principle of 
relativistic invariance is lost, making it unclear not only how to implement
crossing but even whether such a property should hold at all. 

The aim of this paper is to propose how to implement crossing
properties directly in the context of the S-matrix (\ref{e.sfull}). We
believe that on general grounds it is very probable that some form of
crossing symmetry should hold for the $AdS_5 \times S^5$
superstring. Firstly, if one would attempt quantization without using
light-cone gauge, but at the cost of introducing ghosts it is probable
that one could get an explicitly relativistic QFT, which would thus
have to obey crossing conditions. Secondly, there are subsectors of
the full  $AdS_5 \times S^5$ worldsheet theory, which can be
formulated in an explicitly relativistic manner \cite{gleb2}. There
one also expects crossing to hold. For these reasons we believe
that a form of the crossing condition should exist for
(\ref{e.sfull}). In order to find its concrete form we will use an
algebraic formulation of this condition which 
appears in the language of Hopf algebras, the natural
mathematical framework for incorporating nonlocal symmetries in
integrable quantum field theories and which includes as notable
examples the Yangian and quantum affine algebras (see
e.g. \cite{ChariP,Mckay,bernardY}). Applying this
procedure in the context of the S-matrix (\ref{e.sfull}) we find that
the form of crossing transformation is uniquely fixed. We then find an
analog of the rapidity parametrization in relativistic QFT involving a
coupling constant dependent elliptic curve, and use
this to derive equations for the scalar factor $S_0(z_1,z_2)$, where
$z_i$ are the `generalized' rapidities.

The plan of this paper is as follows. In section 2 we recall the
status of crossing symmetry in relativistic integrable quantum field
theories, and we emphasize its Hopf-algebraic reformulation which will
be the basis of our construction. In section 3 we describe in some
detail the construction of the $su(2|2)$ S-matrix of \cite{beisert22}
in the form which will be convenient for our purposes. In section 4 we
derive the form of the crossing transformation and exhibit crossing
properties of the S-matrix. In section 5 we proceed to derive a
generalized rapidity parametrization in the context of (\ref{e.sfull})
which directly generalizes the rapidity variable $\th$ of a
relativistic QFT. We use that to write, in section 6, the unitarity
and crossing equations for the scalar factor $S_0(z_1,z_2)$ in the
generalized rapidity plane. We close the paper with some conclusions
and two appendices.

\section{Crossing symmetry in relativistic integrable QFT}

In a relativistic integrable QFT with some symmetry group the S-matrix
has the form
\eq
S(\th_1,\th_2)=S_0(\th_1-\th_2) \cdot \hat{S}(\th_1-\th_2)
\eqx 
where the $\th_i$ are the {\em rapidities} which parametrize the
energies and momenta of the particles through
\eq
\label{e.rapidity}
E_i=m \cosh \th_i \qqqq p_i=m \sinh \th_i
\eqx
Relativistic invariance ensures that the S-matrix is a function only of
the difference $\th=\th_1-\th_2$. The matrix $\hat{S}(\th)$ is
typically uniquely fixed, up to mutiplication by a scalar function
$S_0(\th)$, using either the Yang-Baxter equation or nonlocal
symmetries from Yangian or quantum affine algebras.

Subsequently the scalar factor is fixed by requiring unitarity
\eq
S^{nm}_{ij}(\th) S^{kl}_{nm}(-\th)=\dl^k_i \dl^l_j
\eqx
and crossing invariance of the S-matrix
\eq
\label{e.relcross}
S^{l\bar{i}}_{\bar{k}j} (i\pi-\th) = S^{kl}_{ij} (\th)
\eqx
where the bars over indices indicate a change from a particle to an
antiparticle and may typically involve some nontrivial action of a
charge conjugation matrix.

For the case of the $su(2|2)\times su(2|2)$ symmetric S-matrix
relevant for the $AdS_5 \times S^5$ superstring it is
unclear how to generalize the crossing relation written in the form
(\ref{e.relcross}). Firstly, the S-matrix depends nontrivially on two
variables and cannot be written as a function of a single variable.
Secondly, one does not know how to implement charge conjugation and
even more importantly what is the analogue of the $i\pi-\th$ in
(\ref{e.relcross}). In order to overcome these difficulties we use a
reformulation of the crossing condition in terms of an underlying
symmetry algebra which is a Hopf algebra. This is in fact a natural setting
for the symmetry algebras of integrable relativistic QFT as {\em
  nonlocal} symmetry charges are naturally incorporated in this
framework through a nontrivial coproduct (i.e. a prescription of how
the nonlocal charge acts on a 2-particle state). The relativistic
crossing symmetry requirement has been already translated into this
framework (see \cite{delius,bernard} and \cite{leclair} for the
supersymmetric case) using another ingredient of a
Hopf algebra -- the antipode.

In order to motivate this formulation let us rewrite the relation
(\ref{e.relcross}) reintroducing two separate rapidities $\th_1$ and
$\th_2$ and keeping in mind that crossing involves changing the first
particle into an antiparticle. Equation (\ref{e.relcross}) can then be
rewritten as
\eq
S(i\pi-\th_1+\th_2)^{cross}=S(\th_1-\th_2)
\eqx
where the superscript ${}^{cross}$ stands for the relevant
transformation of the indices. Now we reverse the signs of $\th_1$ and
$\th_2$ to get
\eq
\label{e.crossref}
S(\th_1+i\pi -\th_2)^{cross}=S(-(\th_1-\th_2))=S(\th_1-\th_2)^{-1}
\eqx
where in the last equality we used unitarity. 
We see that the
particle to antiparticle transformation involves a shift of the
rapidity $\th \to \th+i\pi$ which reverses the signs of the energy and
momentum.

It is this last form
which has a direct Hopf-algebraic interpretation. To see that let us introduce
the R-matrix $R(\th_1,\th_2)$ related to the S-matrix through a
(graded) permutation $P$:
\eq
S(\th_1,\th_2)=P R(\th_1,\th_2)
\eqx
The R-matrix in a Hopf algebra satisfies the direct counterpart of
(\ref{e.crossref}) (see e.g. proposition 4.2.7 in \cite{ChariP}) 
\eq
\label{e.hopfcross}
(\ant \otimes id) R=R^{-1} \qqqq (id \otimes \ant^{-1})R=R^{-1}
\eqx
where $\ant$ is the antipode mapping which has the physical
interpretation of a particle to antiparticle transformation. Directly
from the Yangian or quantum affine algebra viewpoint one shows that
the antipode involves a shift of $\th$ (since in this framework the
rapidity labels the representations of the Yangian), and also possibly
some charge conjugation matrices when the above equations are
considered in some definite representation of a Lie (super)algebra.

We would like to emphasize that (\ref{e.hopfcross}) does not involve
any assumptions on the underlying relativistic invariance of the
theory. The action of the antipode follows just purely
algebraically from the relevant Hopf algebra. Due to this property, we
propose to use (\ref{e.hopfcross}) as a basis for generalizing
crossing to the case of the worldsheet $AdS_5 \times S^5$ theory. We
will perform this construction in section~4.

Before we end this section let us discuss the rapidity parametrization
(\ref{e.rapidity}). It was introduced in order for the $S$ matrix to
be a meromorphic (single-valued) function without any cuts which would
appear if the $S$ matrix was considered as a function of physical
momenta. Since we do not have much intuition about the analytical
structure of the S-matrix for the nonstandard dispersion relations
characteristic of the $AdS_5 \times S^5$ worldsheet theory we would
like to abstract the above mentioned characteristic of the rapidity
parametrization and use it as a guiding principle in our case.

The physical relativistic energies and momenta are
linked by the mass-shell condition
\eq
E^2-p^2=m^2
\eqx 
The rapidity parametrization can be thought of as a uniformization of
the above curve (roughly as a universal covering space\footnote{Note,
  however, that the universal cover here is the sphere and the
  rapidity parametrization can be considered as a mapping onto that
  sphere from the plane (albeit with an essential singularity at
  infinity).}) -- namely a mapping by single-valued functions
from the rapidity plane of complex $\th$. 
\eq
E=m \cosh \th \qqqq p=m \sinh \th
\eqx
The particle-antiparticle
interchange $(E,p) \to (-E,-p)$ then becomes a translation $\th\to
\th+ i\pi$ which is no longer an involution, a fact that will be
important for us later. We propose to implement the crossing
conditions (\ref{e.hopfcross}) for the $AdS_5 \times S^5$ worldsheet
S-matrix on the appropriate universal covering space which would play
the role of the space of generalized rapidities, and by its very definition
would avoid the appearance of cuts. We will construct the universal
covering space in this context in section~5 and write the resulting crossing
equations for the scalar factor in section~6.

\section{The $su(2|2)$ S-matrix}

In \cite{beisert22} the S-matrix has been considered mainly with
explicit (centrally-extended) $su(2|2)$ symmetry. This was at the cost
of introducing explicit length changing operators which have
nontrivial braiding (commutation) relations with the excitations. A
consequence of that was the fact that these braiding factors had to be
incorporated when verifying the Yang-Baxter equation. This nonstandard 
modification of the Yang-Baxter equation would make the Hopf-algebraic
interpretation advocated in the present paper quite
problematic. However, as pointed out in appendix B of
\cite{beisert22}, one can formulate the scattering matrix with only a
manifest $su(1|2)$ symmetry, but with the length changing operators
eliminated. Then the Yang-Baxter equation is just the ordinary YBE
wihout any braiding factors. For these reasons we will adopt here the
$su(1|2)$ symmetric formulation. For completeness we will now
recapitulate in some detail the derivation of the $su(2|2)$ symmetric
S-matrix in this formulation as we will use some of the explicit
constructions presented here in the derivation of the crossing
relations.   

A word of caution is necessary here concerning the identification of
the $su(2|2)\times su(2|2)$ S-matrix of \cite{beisert22} with the
S-matrix of the worldsheet $AdS_5 \times S^5$ superstring. In the
latter paper the S-matrix was formulated from the spin chain point of
view, but in fact all arguments could be recast as implementing
the symmetries in a Zamolodchikov-Faddeev algebra for the worldsheet
theory extended by the length-changing operators. One cannot be
completely certain that this is the only way of implementing these
symmetries but the high degree of uniqueness of the resulting S-matrix
makes it quite probable that this is indeed so. In this paper we will
therefore adopt this hypothesis for the S-matrix of the worldsheet
theory.   

We will denote, as in \cite{beisert22}, the basis states as
$\ket{\phi}\equiv \ket{\phi^1}$, $\ket{\chi} \equiv \ket{\phi^2{\cal Z}^+}$ and
the fermionic $\ket{\psi^{1,2}}$. The operators which will form the
$su(1|2)$ subalgebra have the bosonic index restricted to 1. For
completeness let us quote the action of the generators of $su(1|2)$ on
this basis states. The bosonic rotation generators are canonical:
\eqn
\RR&=&\f{1}{2} \ket{\phi}\bra{\phi}-\f{1}{2} \ket{\chi}\bra{\chi} \\
\LL^\al_\bt &=& \dl_{\gm\bt} \ket{\psi^\al}\bra{\psi^\gm}- \f{1}{2}
\dl^\al_\bt \ket{\psi^\gm}\bra{\psi^\gm}
\eqnx
The fermionic supercharges act on the basis states as follows 
\eqn
\QQ^1=a \ket{\psi^1}\bra{\phi}+b \ket{\eta}\bra{\psi^2} &\qqq&
\QQ^2=a \ket{\psi^2}\bra{\phi}-b \ket{\eta}\bra{\psi^1} \\
\SS_1=c \ket{\psi^2}\bra{\chi}+d \ket{\phi}\bra{\psi^1} &\qqq&
\SS_2=-c \ket{\psi^1}\bra{\chi}+d \ket{\phi}\bra{\psi^2}
\eqnx
The (complex) parameters $a$, $b$, $c$, $d$ parametrize the allowed
representations and encode the energy and momenta of the states. They
are not completely unconstrained. The commutation relation
\eq
\{\QQ^\al, \SS_\bt\} =\LL^\al_\bt +\dl^\al_\bt \RR -\dl^\al_\bt \HH
\eqx
leads to the relation
\eq
\label{e.det}
ad-bc=1
\eqx
while the central charge $\HH$ is fixed to be
\eq
\HH=\left( \f{1}{2} +bc \right) \cdot id
\eqx
From the point of view of the full $psu(2,2|4)$ symmetry of the $AdS_5 \times
S^5$ superstring, $\HH$ is related to the anomalous dimension $\Delta$
through
\eq
\HH=\f{1}{2} \left( \Delta-J \right)
\eqx
Moreover $a$ can be absorbed into relative normalization of the
fermionic and bosonic states. We will set it usually to 1.

\subsubsection*{The $su(1|2)$ invariant S-matrix}

The $su(1|2)$ invariant S-matrix acts in the tensor product of two
representations as 
\eq
S:V_1 \otimes V_2 \to V_2 \otimes V_1
\eqx 
It will turn out to be more convenient to consider the R matrix acting
as
\eq 
R:V_1 \otimes V_2 \to V_1 \otimes V_2
\eqx 
related to the S matrix through the graded permutation operator
$\flip$:
\eq
S=\flip R
\eqx
From $su(1|2)$ invariance the R matrix is then a linear combination of
projectors onto the three irreducible $su(1|2)$ representations
appearing in the tensor product $V_1 \otimes V_2$:
\eq
\label{e.rmat12}
R=S_1\cdot \proj_1 +S_2\cdot \proj_2 +S_3\cdot \proj_3
\eqx
For our purposes we will need an explicit construction of these
projectors. To this end we may first construct the $su(1|2)$ invariant
Casimir operator in $V_1 \otimes V_2$:
\eq
\cas=\f{1}{2} \left( \QQ^\al \SS_\al -\SS_\al \QQ^\al \right)
+(\RR+\HH)^2 -\f{1}{2} \LL^\al_\bt \LL^\bt_\al
\eqx
This operator has three distinct eigenvalues in  $V_1 \otimes V_2$
corresponding to the three irreducible representations. These are
\eqn
\lm_1 &=& (2+b_1c_1+b_2c_2)^2-(b_1c_1+b_2c_2)-2 \\
\lm_2 &=& (1+b_1c_1+b_2c_2)^2-1 \\
\lm_3 &=& (b_1c_1+b_2c_2)(1+b_1c_1+b_2c_2) 
\eqnx
The projectors can then be constructed explicitly as e.g.
\eq
\proj_1=\f{(\cas-\lm_2)(\cas-\lm_3)}{(\lm_1-\lm_2)(\lm_1-\lm_3)}
\eqx
and similar formulas for $\proj_2$ and $\proj_3$.

\subsubsection*{Implementation of the $su(2|2)$ symmetry}

It turns out \cite{beisert22} that it is necessary to extend $su(2|2)$
by additional central charges which however act in such a way that
they vanish on states with vanishing {\em total} momentum i.e. exactly
the states that satisfy `level-matching' for the closed
string\footnote{Or cyclic traces on the gauge theory/spin-chain
  side.}. The two conditions which arise link the $a,b,c,d$ parameters
with the physical momentum $p$:
\eq
ab=\al (e^{-ip}-1) \qqqq cd=\bt (e^{ip}-1)
\eqx
Where $\al$ and $\bt$ are constants {\em common} for all
excitations. Let us note that the above two equations lead to an
additional constraint linking $a,b,c,d$. Namely calculating $e^{ip}$
from one of these equations and inserting into the other one leads to
a quartic constraint
\eq
\label{e.quartic}
abcd+ \bt ab +\al cd=0
\eqx 
The coupling constant is linked to $\al$ and $\bt$ through
$\al\bt=g^2/2$. This condition restricts further the allowed parameters
labeling the explicit $su(1|2)$ representations.

Let us also quote a convenient parametrization of $a,b,c,d$ in terms
of the $x^\pm$ parameters \cite{beisert22} (with $a=1$):
\eq
\label{e.xplmndef}
a=1 \qqq b=-\al \left(1-\f{x^-}{x^+}\right) \qqq c=\f{i\bt}{x^-} \qqq
d=-i(x^+-x^-)
\eqx
where the parameters $x^\pm$ satisfy
\eq
\label{e.xconstr}
x^+ + \f{\al\bt}{x^+} -x^- -\f{\al\bt}{x^-}=i
\eqx 
The inverse mapping is given by
\eq
\label{e.xplmn}
x^-=\f{i\bt}{c} \qqqq x^+=\f{\al x^-}{\al+b}
\eqx

In order to completely determine the scalar
coefficients $S_i$ in (\ref{e.rmat12}) it is enough to assume symmetry
under one of the other $su(2|2)$ supercharges (see
\cite{beisert22}). In the explicit $su(1|2)$ setup these supercharges
add the length-changing operators ${\cal Z}$ or ${\cal
  Z}^+$ to {\em all} states in the multiplet. Therefore invoking these
symmetries will necessarily involve braiding factors. Explicitly the
S-matrix has to satisfy
\eq
(B_1\, \tilde{\QQ}_2 \otimes id+(-1)^F \otimes \tilde{\QQ}_1) \cdot S -
S \cdot (B_2\, \tilde{\QQ}_1 \otimes id+(-1)^F \otimes \tilde{\QQ}_2)=0
\eqx
where $B_i$ are braiding factors coming from the commutation of
length-changing operators. It turns out that the above equation for
just one of the supercharges 
\eq
\tilde{\QQ}_i=a_i \ket{\psi^1}\bra{\chi}-b_i\ket{\phi}\bra{\psi^2}
\eqx
fixes uniquely both the braiding factors $B_i$ and the S-matrix
coefficients $S_i$ up to a common scalar factor. In terms of the
$a,b,c,d$ the expressions are quite lengthy but become rather simple in terms
of $x^{\pm}$. We recover thus the result of Appendix B of
\cite{beisert22}:
\eqn
S_1&=&\f{x_2^+-x_1^-}{x^-_2-x_1^+} \\
S_2&=& 1 \\
S_3&=& \f{x_2^-}{x_2^+} \f{x_1^+}{x_1^-}  \f{x_2^+-x_1^-}{x^-_2-x_1^+}
\\
B_i &=& \f{x_i^-}{x_i^+}
\eqnx
where $S_2=1$ was chosen as in \cite{beisert22} to normalize the
S-matrix to the gauge theory asymptotic S-matrix.

\subsubsection*{The full $su(2|2)\times su(2|2)$ S-matrix}

The full $su(2|2)\times su(2|2)$ S-matrix is uniqely fixed by imposing
$su(2|2)$ symmetry in each factor. The two algebras are linked by
sharing the same $\HH$ operator. Therefore one is led to use the same
parameters $x^\pm$ (or $a,b,c,d$) for both factors. One therefore has
essentially 
\eq
S(p_1,p_2)=S_0(p_1,p_2) \cdot \left[ \hat{S}_{su(2|2)}(p_1,p_2) \otimes
\hat{S}_{su(2|2)}(p_1,p_2) \right]
\eqx
However care has to be taken to implement correctly all
anticommutation relations which necessitates the introduction of various
factors of $(-1)^F$ -- therefore the tensor product notation used
above has to be understood somewhat symbolically. Let us note that
one may also use a different explicit algebra for one of the $su(2|2)$
factors which would also eliminate length-changing processes
i.e. $su(2|1)$ instead of $su(1|2)$. We checked that this does not
modify the crossing properties derived in the following section apart
from a trivial `braiding factor' so we will not consider further this
possibility.

\section{The antipode and crossing properties}

In this section we will implement the formulation of crossing property
of the R-matrix using the Hopf-algebraic conditions
\eq
\label{e.SReq}
(\ant \otimes id)R=R^{-1} \qqqq (id \otimes \ant^{-1}) R=R^{-1}
\eqx
where $\ant$ is the antipode. It is an antihomomorphism of the Hopf
(super)-algebra i.e.
\eq
\ant(AB)=(-1)^{d(A)d(B)} \ant(B)\ant(A)
\eqx
The above equations descend to equations in specific representations
$V_1 \otimes V_2$ through the introduction of a charge conjugation
matrix $\chrg$:
\eq
\label{e.antipode}
\pi(\ant(A))= \chrg^{-1} \bar{\pi}(A)^{st} \chrg
\eqx
where $\bar{\pi}$ is the representation for the antiparticles which
can be distinct from $\pi$, and ${}^{st}$ stands for the
supertranspose defined as
\eq
M^{st}_{ij}=(-1)^{d(i)d(j)+d(j)}\, M_{ji}
\eqx

In order to apply the above framework to the case of the $su(2|2)$
(and consequently the full $su(2|2)\times su(2|2)$) S-matrix we
encounter some difficulties. First we do not have a complete
description of the Hopf algebra of (nonlocal) symmetries of the
S-matrix. There are strong indications that it is not in fact a
Yangian. However we will assume that the explicit $su(1|2)$ algebra is
part of the full Hopf algebra, and implementing (\ref{e.antipode}) for
$A\in su(1|2)$ will determine $\chrg$ and partially the representation
$\bar{\pi}$. Then we will find that in order for (\ref{e.SReq}) to
have any chance of having a solution will uniquely determine the remaining
freedom in $\bar{\pi}$ and give equations for the scalar factor $S_0$.

For algebra elements belonging to a Lie superalgebra the antipode acts
very simply as $\ant(A)=-A$. Then the equation (\ref{e.antipode})
takes the form
\eq
\label{e.sacom}
\chrg \cdot \pi(A) + \bar{\pi}(A)^{st} \chrg=0
\eqx
Let us assume that the representation $\pi$ is defined through the
parameters $a_1,b_1,c_1$ and $d_1$. 
Inserting the generators of $su(1|2)$ into (\ref{e.sacom}), we obtain
one constraint on the representation $\bar{\pi}$
\eq
\label{e.barc}
\bar{c}=-\f{1+b_1 c_1}{\bar{b}}
\eqx
The charge conjugation matrix $\chrg$ is then seen to act as follows
on the basis states
\eqn
\chrg \ket{\phi}=\f{a_1 b_1}{\bar{a}\bar{b}} \ket{\chi} &\qqqq&
\chrg \ket{\chi}=\ket{\phi} \\
\chrg \ket{\psi^1}=-\f{b_1}{\bar{a}} \ket{\psi^2} &\qqqq& 
\chrg \ket{\psi^2}=\f{b_1}{\bar{a}} \ket{\psi^1}
\eqnx
where an overall factor is arbitrary but in any case it cancels out
from all subsequent equations.
The equation (\ref{e.sacom}) does not lead to any constraints on
$\bar{b}$ (as $\bar{a}$ is again just an arbitrary normalization). We believe
that if we would know the full Hopf algebra and therefore the action of the
antipode on the nonlocal generators, we could also determine
$\bar{b}$ directly from (\ref{e.antipode}). We will determine it
however using other reasoning. 

Let us consider the first equation in (\ref{e.SReq}) rewritten using
(\ref{e.antipode}) and denoting the R matrix with the proper scalar
factor by $R_{fin}$: 
\eq
\label{e.rcond}
(\chrg^{-1} \otimes id)\, R_{fin}(\bar{1},2)^{st_1}\,  (\chrg \otimes id) \, 
R_{fin}(1,2)=id
\eqx
where ${}^{st_1}$ denotes the supertranspose in the first entry of
$R_{fin}(\bar{1},2)$ defined explicitly as
\eq
\left(R^{st_1}\right)^{b_1 b_2}_{a_1a_2} =(-1)^{d(a_1)d(b_1)+d(a_1)}
R^{a_1b_2}_{b_1a_2} 
\eqx
Typically given a solution of the Yang-Baxter equation $R(1,2)$ which
is invariant 
under all relevant symmetries fixes $R_{fin}(1,2)$ up to scalar
multiplication by a function $S_0(1,2)$:
\eq
\label{e.rfin}
R_{fin}(1,2) \equiv S_0(1,2) R(1,2)
\eqx
In our case $R(1,2)$ is the solution (\ref{e.rmat12}). Inserting
(\ref{e.rfin}) into (\ref{e.rcond}) we find that in order for the
S-matrix to be crossing symmetric the expression
\eq
\label{e.ident}
(\chrg^{-1} \otimes id)\, R(\bar{1},2)^{st_1}\,  (\chrg \otimes id) \, 
R(1,2)
\eqx  
has to be a multiple of the identity. This is a very nontrivial
equation which {\em a-priori} does not need to hold at all. 

In order to analyze it in detail we first determine how do the individual
representations transform under crossing:
\eq
\label{e.projcross}
(\chrg^{-1} \otimes id)\, \proj_i(\bar{1},2)^{st_1}\,  (\chrg \otimes
id)= M_{ik} \, \proj_k(1,2)
\eqx 
It turns out that the matrix $M_{ik}$ is quite nontrivial and does
not have any vanishing entries (see the explicit formulas in
Appendix~B). Now the requirement that
(\ref{e.ident}) equals\footnote{We write the scalar function in this
  form for later convenience.} $1/f(1,2) \cdot id$ is equivalent to the system
of three scalar equations:
\eqn
\label{e.eqf1}
S_i(\bar{1},2)\, M_{i1}\, S_1(1,2) &=& 1/f(1,2) \\
S_i(\bar{1},2)\, M_{i2}\, S_2(1,2) &=& 1/f(1,2) \\
\label{e.eqf3}
S_i(\bar{1},2)\, M_{i3}\, S_3(1,2) &=& 1/f(1,2)
\eqnx
Equating the left hand sides of the first two equations gives two
solutions for $\bar{b}$. Equating subsequently the left hand side of
the third equation picks a unique choice for $\bar{b}$:
\eq
\bar{b}=\f{a_2 b_1 b_2(a_1 c_2(1+b_2 c_2)-a_2 c_1 (1+b_1 c_1)}{
  \bar{a} c_2 (a_1b_1-a_2b_2)(1+b_2 c_2)}
\eqx
Now this expression should be a function only of $a_1,b_1,c_1$ and
should {\em not} depend on the second particle. Quite remarkably, one
can show using (\ref{e.quartic}) that the dependence on the second
particle cancels out and $\bar{b}$ becomes
\eq
\label{e.barb}
\bar{b}=\f{-\al a_1 b_1}{\bar{a} (\al+a_1 b_1)}
\eqx
which fixes uniquely\footnote{Up to a trivial rescaling of $\bar{a}$
  which we will set to 1.} the representation $\bar{\pi}$. We believe,
that the fact that such a solution depending only on $a_1,b_1,c_1$
exists at all, is a very strong indication that crossing symmetry
should hold for the $AdS_5 \times S^5$ worldsheet theory.

\subsubsection*{The crossing transformation}

Before we present the result for the scalar factor $f(1,2)$, let us
examine more closely the interpretation of the particle-antiparticle
transformation (\ref{e.barc}) and (\ref{e.barb}). Expressing the
transformation of the $x^{\pm}$ induced by these formulas we get the
very simple result
\eq
\label{e.xcross}
x^+ \to \f{\al \bt}{x^+} \qqqq x^- \to \f{\al\bt}{x^-}
\eqx
Using $e^{ip}=x^+/x^-$ we see that the momentum changes sign. The same
also holds for the energy. This is reassuring since the analogous
transformation for a relativistic theory $\th \to \th+i\pi$ also
reverses the signs of both the momentum and energy. Let us note,
however, that we did not assume the form of transformation
(\ref{e.xcross}) but obtained it purely algebraically.

We may now obtain the function $f(1,2)$ from any of the equations
(\ref{e.eqf1})-(\ref{e.eqf3}). Again the expression in terms of the
$a_i,b_i$ and $c_i$ is quite complicated but it simplifies
considerably when expressed in terms of the $x_i^\pm$ variables:
\eq
\label{e.f}
f(1,2)=\f{\left(\f{\al\bt}{x_1^+} -x_2^-\right) \left(x_1^+-x_2^+
  \right)}{ \left(\f{\al\bt}{x_1^-} -x_2^-\right) \left(x_1^--x_2^+
  \right)} 
\eqx
Let us first note that the above function is not a constant, so a
nontrivial scalar factor is needed in order to form a crossing
symmetric S-matrix. If we would be interested only in a $su(2|2)$
symmetric S-matrix we would thus be led to the equation
\eq
S_0(\bar{1},2) S_0(1,2)=f(1,2)
\eqx
On the other hand if we consider the case of our main interest
i.e. $su(2|2) \times su(2|2)$ symmetry the relevant equation would be
\eq
\label{e.cross1}
S_0(\bar{1},2) S_0(1,2)=f(1,2)^2
\eqx
We have verified the above by explicitly constructing the
$su(2|2)\times su(2|2)$ symmetric S-matrix taking into account various
$(-1)^F$ factors and also the various signs appearing in the
supertranspose. We found that the relevant `crossing' function is
indeed just the square of $f(1,2)$.  

The above equation (\ref{e.cross1}) has to be supplemented by 
the unitarity equation 
\eq
S_0(1,2) S_0(2,1)=1
\eqx
and an analogous crossing relation for the second particle (i.e. $2 \to
\bar{2}$) which is obtained using
\eq
\label{e.rcond2}
(id \otimes \chrg^{-1})\, R_{fin}(1,\bar{2})^{st_2}\,  (id \otimes \chrg) \, 
R_{fin}(1,2)=id
\eqx
where ${}^{st_2}$ is defined as
\eq
\left(R^{st_2}\right)^{b_1 b_2}_{a_1a_2} =(-1)^{d(a_2)d(b_2)+d(b_2)}
R^{b_1a_2}_{a_1b_2} 
\eqx
The result is
\eq
S_0(1,\bar{2})S_0(1,2)=f(1,2)^2
\eqx

Let us note one slightly troubling property of (\ref{e.cross1}). The
right hand side is {\em not} symmetric under the interchange
(\ref{e.xcross}), while the left hand side apparently is. This would
lead to an apparent contradiction. 
In order to avoid that conclusion we have to keep in mind that $x^\pm$
are not independent 
variables but are linked through the constraint (\ref{e.xconstr}). So
such an expression as (\ref{e.f}) has cuts. As advocated in section~2,
in order to deal with
meromorphic functions we would have to pass to the universal covering
space of (\ref{e.xconstr}) or equations (\ref{e.det}) and
(\ref{e.quartic}) and only there one would have to consider the equation
(\ref{e.cross1}). On this covering space the transformation $1 \to
\bar{1}$ would no longer necessarily be an involution.

Such behaviour in fact holds for the conventional case of the
rapidity parametrization in relativistic QFT. There the $1 \to
\bar{1}$ transformation corresponds to $\th \to \th+i\pi$ which does
not square to the identity (while the original transformation
$(E,p)\to (-E,-p)$ on the curve $E^2-p^2=m^2$ is an involution). Also
for such generic Hopf algebras involving nonlocal symmetry charges
like Yangians, the square of the antipode is not equal to the identity
$\ant^2 \neq id$.  Therefore we expect the universal cover to be the
natural algebraic scene for considering the S-matrices. Of course we
have to resort to such algebraic arguments since we lack a deeper
physical understanding of the structure of the worldsheet theory.

In the next section we will explicitly construct the universal
covering space of the parameter space given by (\ref{e.det}) and
(\ref{e.quartic}), and in section 6 we will write the crossing and
unitarity equations directly on that generalized `rapidity plane'.

\section{The generalized rapidity plane}

In this section we would like to introduce analogues of the rapidity
variable $\th$ in relativistic quantum field theory. Let us recall
that, as described in section~2, the rapidity parameter space can be
understood to be 
the universal cover\footnote{With the caveat of an additional mapping
  from the plane as mentioned above.}  of the $(E,p)$
variables subject to the constraint of the relativistic mass shell
condition $E^2-p^2=m^2$. Intuitively this means that the rapidity
variable $\th$ allows to get rid of the cuts and to deal with purely
meromorphic functions. 

The $su(2|2)$ S-matrix is expressed in terms of the complex parameters
$a,b,c,d$ of the $\QQ^\al$ and $\SS_\al$ operators  subject to two
constraints\footnote{A similar analysis directly in terms of the $x^+$
  and $x^-$ parameters will be performed in Appendix~A.}:
\eqn
ad-bc &=& 1\\
abcd+\bt ab+\al cd &=& 0
\eqnx   
where the first constraint is necessary for $sl(1|2)$ symmetry, while
the second quartic constant follows from the structure of the central
charges in order to have $su(2|2)$ symmetry at zero total
momentum. The $a$ parameter can be absorbed in the relative
normalization of the states and so we will set it to 1. Consequently
one can express $d$ as $d=1+bc$ and the remaining {\em nonlinear}
constraint is
\eq
\label{e.curve}
(bc^2+c)(b+\al)+\bt b=0
\eqx
The above algebraic curve has degree 4, and so if it would be
nonsingular it would have genus $g=(4-1)(4-2)/2=3$. It turns out,
however, that it contains singularities: a double point and a cusp
which brings down the genus to 1. This means that it can be
uniformized by elliptic functions defined on the complex plane, which
is the universal cover of (\ref{e.curve}). 

We will now find the explicit form of these mappings.
First let us express $c$ as
\eq
c=\f{-1+v}{2b}
\eqx
where $v$ satisfies the equation
\eq
v^2=1-4\f{\bt b^2}{b+\al}
\eqx
multiplying both sides by $\bt^2 (b+\al)^2$, and introducing $y\equiv
\bt(b+\al)v$ we obtain
\eq
y^2=-4\bt^3 b^3+(1-4\al\bt)\bt^2 b^2+2\al \bt^2 b+\al^2 \bt^2
\eqx
Now a final coordinate transformation $b=-(x+(4\al\bt-1)/12)/\bt$
reduces this elliptic curve to the standard Weierstrass form
\eq
\label{e.elliptic}
y^2=4x^3-g_2 x-g_3
\eqx
where 
\eqn
g_2 &=&\f{1}{12}(1+16\al\bt +16\al^2\bt^2) \\ 
g_3 &=&\f{1}{216} (1+8\al\bt)(-1-16\al\bt+8\al^2 \bt^2) 
\eqnx
The resulting parametrization is $y=\PP'(z)$ and $x=\PP(z)$. So finally
\eqn
b(z) &=& -\f{1}{\bt} \left(\PP(z)+\f{4\al\bt-1}{12}\right)  \\
c(z) &=& \f{\PP'(z)-\bt b(z)-\al\bt}{2 \bt b(z) (b(z)+\al)}
\eqnx
Using formulas (\ref{e.xplmn}) we may obtain the
functions\footnote{An alternate but essentially equivalent
  parametrization is derived in Appendix A.}
$x^+(z)$ and $x^-(z)$.
Before we proceed let us describe in more detail the elliptic curve
(\ref{e.elliptic}). It depends explicitly on the gauge theory coupling
constant $g^2=2\al\bt$. For nonzero (real) coupling it is nonsingular,
its discriminant is 
\eq
\Delta=g_2^3-27 g_3^2 =\al^4 \bt^4 (1+16\al\bt)
\eqx
It has two half-periods $\om_1$ and $\om_2$ which, together with
$\om_3\equiv -\om_1-\om_2$ get mapped by the Weierstrass function
$\PP(z)$ to the zeroes $e_1,e_2,e_3$ of the polynomial
\eq
4x^3-g_2 x-g_3=4(x-e_1)(x-e_2)(x-e_3)
\eqx
Using the explicit forms of $g_2$ and $g_3$ we find that one of the
zeroes has the following simple form:
\eq
e_1=\f{1+8\al\bt}{12}
\eqx
We will denote the corresponding half-period by $\om_1$ i.e.
\eq
\label{e.om1}
\PP(\om_1)=e_1
\eqx
We also have another identity which will be useful later
\eq
(e_1-e_2)(e_1-e_3)=\al^2\bt^2
\eqx

\subsubsection*{Crossing transformations in the $z$ plane}

Now we have to see how the particle-antiparticle transformation 
\eq
x^{\pm} \to \f{\al\bt}{x^{\pm}}
\eqx
is represented on the generalized rapidity $z$ plane. It turns out
that this transformation has a very simple representation very similar
to the transformation $\th \to \th+i\pi$ in the relativistic
case.

Using the addition laws\footnote{See e.g. section 20.33 of
  \cite{Whittaker}.} for $\PP(z)$ and $\PP'(z)$
\eq
\PP(z+\om_1)=e_1+\f{(e_1-e_2)(e_1-e_3)}{\PP(z)-e_1}
\eqx 
and
\eq
\PP'(z+\om_1)=-(e_1-e_2)(e_1-e_3) \f{\PP'(z)}{(\PP(z)-e_1)^2}
\eqx
after straightforward but slightly tedious calculations we obtain
\eq
b(z+\om_1)=\bar{b}(z) \qqqq c(z+\om_1)=\bar{c}(z)
\eqx
Consequently we have that
\eq
x^{\pm}(z+\om_1) = \f{\al\bt}{x^{\pm}(z)}
\eqx
Therefore the $1\to \bar{1}$ transformation on the universal cover is
represented by a translation by the half-period $\om_1$ defined
through (\ref{e.om1}):
\eq
z \to z+\om_1
\eqx

There is another natural transformation $x^+ \to -x^-$, $x^- \to -x^+$
which interchanges the sign of momentum while keeping the energy
unchanged. This is represented also very simply by $z \to -z$. 

\vfill
\pagebreak

\section{Crossing equations for the $su(2|2)\times su(2|2)$ S-matrix}

We are now ready to write the final form of the functional equations
for the scalar factor of the $su(2|2)\times su(2|2)$ invariant
S-matrix. We consider it to be defined as a meromorphic function on
two copies of the complex plane which represent the `generalized
rapidities' of the two particles. It has to satisfy the unitarity
equation
\eq
\label{e.unitz}
S_0(z_1,z_2)S_0(z_2,z_1)=1
\eqx
and crossing w.r.t. the first and second particle
\eqn
\label{e.cross1z}
S_0(z_1+\om_1,z_2) S_0(z_1,z_2)&=&f(z_1,z_2)^2 \\
\label{e.cross2z}
S_0(z_1,z_2-\om_1) S_0(z_1,z_2)&=&f(z_1,z_2)^2
\eqnx
In the above we used the fact that performing crossing w.r.t to the
second particle leads to the same scalar function $f(1,2)$. The
transformation $z_2 \to z_2-\om_1$ arises from the
Hopf-algebraic crossing equation for the second particle (see
(\ref{e.SReq})) which involves the {\em inverse} of the antipode
$\ant^{-1}$. Let us note that in order for the two crossing conditions
(\ref{e.cross1z})-(\ref{e.cross2z}) to be consistent with unitarity
(\ref{e.unitz}) the function $f(z_1,z_2)$ has to satisfy a nontrivial
consistency relation
\eq
f(z_1-\om_1,z_2) = \f{1}{f(z_2,z_1)}
\eqx
We find that indeed (\ref{e.f}) satisfies the above condition. We
believe that this is another argument for the relevance of such a
crossing condition, formulated on the universal cover, to the $AdS_5
\times S^5$ superstring worldsheet S-matrix. In fact if we would have
chosen the translation $z_2 \to z_2+\om_1$ the resulting equation
would be $f(z_1,z_2)=1/f(z_2,z_1)$ which does {\em not} hold.

The natural question is now to determine a minimal solution
$S_0(z_1,z_2)$ and the corresponding form of CDD factors. The
equations are however quite complicated and the standard iterative
technique for solving the coupled crossing and unitarity relations (see
e.g. \cite{leclair,WSh}) does not work here. We postpone the study of
this issue to a separate publication \cite{RJfuture}.

\section{Conclusions}

In this paper we have proposed how to implement crossing relations for
the $su(2|2)\times su(2|2)$ symmetric S-matrix relevant for the $AdS_5
\times S^5$ superstring worldsheet theory. Once constructed, these
relations provide functional equations for the overall scalar factor
(the so-called `dressing factor') of the S-matrix.

Our proposal involves two basic steps. First, a Hopf-algebraic
reformulation of the relativistic crossing relation allows to address
the problem in a purely algebraic manner. The lack of knowledge of the
full Hopf algebra structure of the nonlocal symmetries allows to
determine only a part of the relations directly from properties of the
antipode, 
however, using the structure of the full S-matrix allowed us to fix
uniquely the remaining ambiguity in the crossing transformation. We found
that the original 
S-matrix (normalized to the asymptotic gauge theory result) transforms
nontrivially under crossing thus necessitating a nonconstant scalar
`dressing factor'.

In a second step, in order to eliminate cuts and to deal only with
meromorphic functions, we proposed to introduce the `generalized rapidity
plane' which is a universal covering space of the space of parameters
appearing in the S-matrix. This space is constructed through a
coupling-constant dependent elliptic curve. On this space the crossing
transformation acts very simply as a translation by a specific
half-period.

Finally we derive functional equations for the scalar `dressing
factor' on the universal covering space. We propose to investigate its
solutions in a forthcoming paper \cite{RJfuture}. Apart from that,
there are numerous interesting directions for further study.
It would be interesting to understand more directly the geometric
structure of the parameter space and perhaps link it more directly to
the properties of the worldsheet theory. Another more mathematical
question would be to try to find the whole structure of the Hopf
algebra relevant in this case and in particular to understand more
intrinsically the mathematical origin of the quartic constraint
(\ref{e.quartic}). Finally it would be also very interesting to make
contact with near-BMN quantization of the $AdS_5 \times S^5$
superstring (like the
very recent work \cite{FPZ}),
especially if formulated in an analogous explicit $su(1|2)$ picture.

\bigskip

\noindent{\bf Acknowledgments.} I would like to thank S{\l}awomir
Cynk, Gerhard G\"{o}tz,
and S{\l}awomir Rams for discussions. This work was supported in part
by the Polish Ministry of Science and Information
Society Technologies grants 2P03B08225 (2003-2006), 1P03B02427
(2004-2007) and 1P03B04029 (2005-2008).

\appendix

\section{Uniformization using $x^{\pm}$ variables}

It is interesting to derive directly the uniformization of the parameters
$x^\pm$ satisfying the defining equations
\eq
\label{e.eq1}
x^+ + \f{\al\bt}{x^+} -x^- -\f{\al\bt}{x^-}=i
\eqx
In particular we should obtain the same elliptic curve as in section
5. Equation (\ref{e.eq1}) can be rewritten as
\eq
x^+-x^-=\f{i x^+ x^-}{x^+ x^- -\al\bt}
\eqx
Denoting $w=-x^+x^-$ and $x^++x^-=iy/(w+\al\bt)$ we may derive the
equation linking $y$ and $w$:
\eq
y^2-w^2=4w (w+\al \bt)^2
\eqx 
performing a final substitution $w=x-(1+8\al\bt)/12$ we obtain finally
the curve in Weierstrass form
\eq
y^2=4 x^3 -g_2 x-g_3
\eqx
which exactly coincides with (\ref{e.elliptic}). Putting all the above
together we obtain finally a parametrization for $x^\pm$:
\eqn
x^+(z) &=& \f{i}{2} \f{\PP'(z)+w(z)}{w(z)+\al\bt} \\
x^-(z) &=& \f{i}{2} \f{\PP'(z)-w(z)}{w(z)+\al\bt}
\eqnx
where
\eq
w(z)=\PP(z)-\f{1+8\al\bt}{12}
\eqx
Let us note the amusing fact that if we compare the above
parametrization and the similar one derived in section 5, the $x^-(z)$
functions coincide, while $x^+(z)$ here corresponds to $\al\bt/x^+(z)$
there. In fact this can be implemented by a linear transformation of
the $z$ variable\footnote{We checked this numerically without,
  however, carrying out an analytic proof.}:
\eq
x^\pm_{section\; 5}(z)=x^\pm_{here}\left(\f{\om_1}{2}+\om_2-z \right)
\eqx
showing that the construction presented here and the one in section 5
are essentially equivalent.

\section{Crossing properties of the elementary projectors}

In this appendix, for completeness we write the elements of the matrix
$M_{ij}$ encoding the transformation properties of the projectors
$\proj_i$ under crossing (see eq. (\ref{e.projcross})):
\eqn
M_{11} &=&
 \frac{{b_1}\,{c_1}\,\left( 1 + {b_1}\,{c_1} \right) }
    {\left( -1 + {b_1}\,{c_1} - {b_2}\,{c_2} \right) \,
      \left( {b_1}\,{c_1} - {b_2}\,{c_2} \right)} \\
M_{12} &=&
   - \frac{{b_1}\,{c_1}\,\left( 1 + {b_2}\,{c_2} \right) }
      {\left( -1 + {b_1}\,{c_1} - {b_2}\,{c_2} \right) \,
        \left( {b_1}\,{c_1} - {b_2}\,{c_2} \right) }  \\
M_{13} &=&
   \frac{{b_2}\,{c_2}\,\left( 1 + {b_2}\,{c_2} \right) }
    {\left( -1 + {b_1}\,{c_1} - {b_2}\,{c_2} \right) \,
      \left( {b_1}\,{c_1} - {b_2}\,{c_2} \right) } \\
M_{21} &=&
   -\frac{2\,\left( 1 + {b_1}\,{c_1} \right) \,
      \left( 1 + {b_2}\,{c_2} \right) }{\left( -1 + {b_1}\,{c_1} - 
        {b_2}\,{c_2} \right) \,
      \left( 1 + {b_1}\,{c_1} - {b_2}\,{c_2} \right) } \\
M_{22} &=&
   \frac{{b_1}\,{c_1}\,\left( 1 + {b_1}\,{c_1} \right)  + 
      {b_2}\,{c_2}\,\left( 1 + {b_2}\,{c_2} \right) }{
      \left( -1 + {b_1}\,{c_1} - {b_2}\,{c_2} \right) \,
      \left( 1 + {b_1}\,{c_1} - {b_2}\,{c_2} \right) } \\
M_{23} &=&
  - \frac{2\,{b_1}\,{b_2}\,{c_1}\,{c_2}}
    {\left( -1 + {b_1}\,{c_1} - {b_2}\,{c_2} \right) \,
      \left( 1 + {b_1}\,{c_1} - {b_2}\,{c_2} \right) } \\
M_{31} &=&
   \frac{{b_2}\,{c_2}\,\left( 1 + {b_2}\,{c_2} \right) }
    {\left( {b_1}\,{c_1} - {b_2}\,{c_2} \right) \,
      \left( 1 + {b_1}\,{c_1} - {b_2}\,{c_2} \right) } \\
M_{32} &=&
   - \frac{{b_2}\,{c_2}\,\left( 1 + {b_1}\,{c_1} \right) }
      {\left( {b_1}\,{c_1} - {b_2}\,{c_2} \right) \,
        \left( 1 + {b_1}\,{c_1} - {b_2}\,{c_2} \right) } \\
M_{33} &=&
   \frac{{b_1}\,{c_1}\,\left( 1 + {b_1}\,{c_1} \right) }
    {\left( {b_1}\,{c_1} - {b_2}\,{c_2} \right) \,
      \left( 1 + {b_1}\,{c_1} -
    {b_2}\,{c_2} \right) }
\eqnx

\end{document}